\documentclass[prd,superscriptaddress,amsfonts,amssymb,amsmath,showpacs,twocolumn]{revtex4-2}
\usepackage{bm}
\usepackage{amsfonts}
\usepackage{latexsym}
\usepackage[latin1]{inputenc}
\usepackage{graphicx}
\usepackage{amsmath}
\usepackage{palatino}
\usepackage{mathpazo}
\usepackage{textcomp}
\usepackage{float}
\usepackage{booktabs}
\usepackage{dcolumn}
\usepackage{hyperref}
\hypersetup{colorlinks,citecolor=blue}
\usepackage{amsmath}
\usepackage{xcolor}
\usepackage{orcidlink}
\usepackage[caption=false]{subfig}
\usepackage{commath}
\def\jnl@style{\it}
\def\aaref@jnl#1{{\jnl@style#1}}

\def\aaref@jnl#1{{\jnl@style#1}}

\def\aj{\aaref@jnl{AJ}}                   
\def\apj{\aaref@jnl{ApJ}}                 
\def\apjl{\aaref@jnl{ApJ}}                
\def\apjs{\aaref@jnl{ApJS}}               
\def\apss{\aaref@jnl{Ap\&SS}}             
\def\aap{\aaref@jnl{A\&A}}                
\def\aapr{\aaref@jnl{A\&A~Rev.}}          
\def\aaps{\aaref@jnl{A\&AS}}              
\def\mnras{\aaref@jnl{Mon.~Not.~Roy.~Astron.~Soc.}}             
\def\prd{\aaref@jnl{Phys.~Rev.~D}}        
\def\prc{\aaref@jnl{Phys.~Rev.~C}}  
\def\prl{\aaref@jnl{Phys.~Rev.~Lett.}}    
\def\qjras{\aaref@jnl{QJRAS}}             
\def\skytel{\aaref@jnl{S\&T}}             
\def\ssr{\aaref@jnl{Space~Sci.~Rev.}}     
\def\zap{\aaref@jnl{ZAp}}                 
\def\nat{\aaref@jnl{Nature}}              
\def\aplett{\aaref@jnl{Astrophys.~Lett.}} 
\def\apspr{\aaref@jnl{Astrophys.~Space~Phys.~Res.}} 
\def\physrep{\aaref@jnl{Phys.~Rep.}}      
\def\physscr{\aaref@jnl{Phys.~Scr}}       
\def\commat{\aaref@jnl{Comm.~Math.~Phys.}}              
\def\science{\aaref@jnl{Science}}               
\def\cqg{\aaref@jnl{Classical Quant.~Grav.}}            
\def\jpcs{\aaref@jnl{JPCS}}                                     
\def\ijmpd{\aaref@jnl{Int.~J.~Mod.~Phys.~D}}                    
\def\grg{\aaref@jnl{Gen.~Relat.~Gravit.}}               
\def\rpp{\aaref@jnl{Rep.~Prog.~Phys.}}          
\def\npa{\aaref@jnl{Nucl.~Phys.~A}}        
\def\lrr{\aaref@jnl{Living Rev.~Rel.}}                   
\def\jcap{\aaref@jnl{J.~Cosmology Astropart.~Phys.}}    
\def\rmp{\aaref@jnl{Rev.~Mod.~Phys.}}   
\def\epjc{\aaref@jnl{Eur.~Phys.~J.~C}}

\allowdisplaybreaks[1]

\begin{document}

\color{black}       

\title{Late-time acceleration in $f\left( Q\right) $ gravity: Analysis and constraints in an anisotropic background}
\author{M. Koussour\orcidlink{0000-0002-4188-0572}}
\email{pr.mouhssine@gmail.com}
\affiliation{Quantum Physics and Magnetism Team, LPMC, Faculty of Science Ben
M'sik,\\
Casablanca Hassan II University,
Morocco.}

\author{K. El Bourakadi\orcidlink{0000-0002-2199-9613}}
\email{elbourkadi17@gmail.com }
\affiliation{Quantum Physics and Magnetism Team, LPMC, Faculty of Science Ben
M'sik,\\
Casablanca Hassan II University,
Morocco.}

\author{S.H. Shekh\orcidlink{0000-0003-4545-1975}}
\email{da\_salim@rediff.com}
\affiliation{Department of Mathematics. S. P. M. Science and Gilani Arts Commerce
College,\\ Ghatanji, Dist. Yavatmal, Maharashtra-445301, India.}

\author{S. K. J. Pacif\orcidlink{0000-0003-0951-414X}}
\email{shibesh.math@gmail.com}
\affiliation{Centre for Cosmology and Science Popularization (CCSP) 
SGT University,\\ Gurugram 122505, Haryana, India.}

\author{M. Bennai\orcidlink{0000-0003-1424-7699}}
\email{mdbennai@yahoo.fr }
\affiliation{Quantum Physics and Magnetism Team, LPMC, Faculty of Science Ben
M'sik,\\
Casablanca Hassan II University,
Morocco.} 
\affiliation{Lab of High Energy Physics, Modeling and Simulations, Faculty of
Science,\\
University Mohammed V-Agdal, Rabat, Morocco.}
%
\date{\today}
\begin{abstract}
This paper is devoted to investigate the anisotropic locally rotationally
symmetric (LRS) Bianchi type-I space-time in the context of the recently
proposed $f(Q)$ gravity in which $Q$ is the non-metricity scalar. For this
purpose, we consider a linear form of $f\left( Q\right) $ gravity model,
specifically, $f\left( Q\right) =\alpha Q+\beta $, where $\alpha $ and $%
\beta $ are free parameters and we analyzed the exact solutions of LRS
Bianchi type-I space-time. The modified Friedmann equations are solved by
presuming an expansion scalar $\theta \left( t\right) $ is proportional to
the shear scalar $\sigma \left( t\right) $ which leads to the relation
between the metric potentials as $A=B^{n}$ where $n$ is an arbitrary
constant. Then we constrain our model parameters with the observational
Hubble datasets of 57 data points. Moreover, we discuss the physical behavior of cosmological
parameters such as energy density, pressure, EoS parameter, and deceleration
parameter. The behavior of the deceleration
parameter predicts a transition from deceleration to
accelerated phases in an expanding Universe. Finally, the EoS parameter
indicates that the anisotropic fluid behaves like the standard $\Lambda $CDM
model.
\end{abstract}

\maketitle

\date{\today}
\section{Introduction}

\label{sec1}

Observations of high redshift supernovae and cosmic microwave background
fluctuations (CMBR) \cite{ref1, ref2, ref3, ref4} indicated that the present
acceleration epoch of the Universe is accelerated. This late-time
acceleration is due to an unidentified fluid called dark energy (DE). Many
suggestions have been considered as a candidate to explain the true nature
of DE. The first is the cosmological constant that encounters problems such
as the incredibly small value required by general relativity (GR) theory. On
other hand, the cosmological constant provided by particle physics
predictions is generally more than 50 orders of magnitude than the actual
value assumed by GR \cite{ref5}.\ This mysterious DE that is responsible for
the cosmological acceleration and is estimated as $68\%$\ \ of the total
energy density of the Universe may require us to reconsider the theory of
gravity on cosmological scales. The DE can be tested using an effective tool
namely the equation of state (EoS) parameter of the form $\omega =\frac{p}{%
\rho }$, which is the ratio of the cosmic pressure $p$ to the cosmic energy
density $\rho $. Each DE model has a different EoS parameter value, for
example, in the case of the cosmological constant mentioned above $\omega
=-1 $, also for the quintessence model $\omega$ is bounded as $-1<\omega
<-0.33$, and finally $\omega <-1$ for the phantom DE model.

Modified gravity theories (MGT) provide intriguing theoretical concepts for
addressing the cosmological constant problem and explaining the late-time
acceleration of the Universe. Several DE models started from the simplest
modified gravity $1/R$ theory \cite{ref6, ref7}. In general, MGT appears to
be quite appealing since it provides subjective solutions to a number of key
problems concerning DE. An alternative theory to GR is teleparallel gravity
by which gravitational interaction is described by the torsion scalar $T$ 
\cite{ref8, ref9, ref10} in a space-time with zero curvature. This theory is
named \emph{teleparallel equivalent to general relativity }(TEGR) and
formulated by tetrad fields on the tangent space in the Weitzenbock
connection which is different from the Levi--Civita connection in GR. The
advantage of working with $f(T)$ models is the order of the field equations,
this allows simplifying the dynamics and finding easily exact solutions.
Symmetric teleparallel $f(Q)$\ gravity is also an alternative theory in
which the covariant derivative of the metric tensor does not vanish, i.e. $%
Q_{\gamma \mu \nu }=\nabla _{\gamma }g_{\mu \nu }$. This theory is called 
\emph{symmetric teleparallel equivalent to general relativity} (STEGR) \cite%
{ref11, ref12}. This new modified $f(Q)$ gravity where $Q$ is the
non-metricity scalar attracted interest of many researchers \cite{ref13,
ref14, ref15, ref16, ref17, ref18}. Moreover, this theory is based on the
generalization of Riemannian geometry described by Weyl geometry \cite{ref19}%
. Generally, the gravitational interaction is classified through three types
of geometries: the curvature of space-time, torsion, and non-metricity. For
this reason, in recent decades researchers have been attracted to MGT
because they reflect the current phenomena of the Universe. Therefore,
gravitational interactions have been calculated using several forms of
geometrics \cite{ref20, ref21, ref22}.

It has been stated by observations that the Universe is homogeneous and
isotropic when the inflationary phase was successfully produced \cite{ref23}%
. However, anomalies in the CMBR lead to conclude that an anisotropic phase
in the early Universe which make it not exactly uniform \cite{ref24}. Thus,
constructing cosmological models that describe the anisotropic and
inhomogeneous properties of the Universe must be taken into consideration.
Toward this goal, Bianchi-type models provide a good description of the
anisotropic background and investigate the cosmic evolution in the early
Universe. In fact, there exist nine types of Bianchi models in the
literature. Here, we consider the anisotropic locally rotationally symmetric
(LRS) Bianchi type-I model which is assumed to be a more general
cosmological metric than Friedmann-Lemaitre-Robertson-Walker (FLRW) metrics 
\cite{ref25}. The Bianchi type-I model is used to test the possible effects
of anisotropy in the early Universe \cite{ref26}. Recently, cosmological
models have been constructed using anisotropic fluid in Bianchi type-I
space-time. Moreover, some exact Bianchi type-I solutions have also been
investigated in $f\left( Q\right) $ modified gravity \cite{ref27, ref28}.
The Bianchi type I model usually presents good consistency with the most
simple mathematical form, considering the nature of this model. Bianchi type
I theory was studied in the context of a viscous fluid to discuss the
behavior of the early Universe near the singularity \cite{ref29}.

The current article is organized as follows: In Sec. \ref{sec2} we discuss
the theoretical basis for $f(Q)$ gravity. In Sec. \ref{sec3}, we derive the
field equations in the LRS Bianchi type-I model. In Sec. \ref{sec4}, the
cosmological solutions of the field equations are calculated with
anisotropic relation. In Sec. \ref{sec5} we analyze the physical and
geometrical parameters of the cosmological model. Further, we constrain our model parameters with the observational
Hubble datasets of 57 data points. Finally, the conclusion of
the results is given in Sec \ref{sec6}.

\section{$f(Q)$ gravity formalism}

\label{sec2}

In differential geometry, the symmetric metric tensor $g_{\mu \nu }$ is used
based on the definition of the length of a vector, and an asymmetric
connection $\Sigma {^{\gamma }}_{\mu \nu }$\ is used to define the covariant
derivatives and parallel transport. Hence, the general affine connection can
be decayed into three components: the Christoffel symbol ${\Gamma ^{\gamma }}%
_{\mu \nu }$, the contortion tensor ${C^{\gamma }}_{\mu \nu }$, and the
disformation tensor ${L^{\gamma }}_{\mu \nu }$, respectively, which is given
by \cite{ref19}

\begin{equation}
\Sigma {^{\gamma }}_{\mu \nu }={\Gamma ^{\gamma }}_{\mu \nu }+{C^{\gamma }}%
_{\mu \nu }+{L^{\gamma }}_{\mu \nu },  \label{eqn1}
\end{equation}%
where the Levi-Civita connection ${\Gamma ^{\gamma }}_{\mu \nu }$ of the
metric $g_{\mu \nu }$ has the form

\begin{equation}
{\Gamma ^{\gamma }}_{\mu \nu }\equiv \frac{1}{2}g^{\gamma \sigma }\left( 
\frac{\partial g_{\sigma \nu }}{\partial x^{\mu }}+\frac{\partial g_{\sigma
\mu }}{\partial x^{\nu }}-\frac{\partial g_{\mu \nu }}{\partial x^{\sigma }}%
\right) ,  \label{eqn2}
\end{equation}%
the contorsion tensor ${C^{\gamma }}_{\mu \nu }$ can be written as

\begin{equation}
{C^{\gamma }}_{\mu \nu }\equiv \frac{1}{2}{T^{\gamma }}_{\mu \nu }+T_{(\mu
}{}^{\gamma }{}_{\nu )},  \label{eqn3}
\end{equation}%
where ${T^{\gamma }}_{\mu \nu }\equiv 2\Sigma {^{\gamma }}_{[\mu \nu ]}$ in
Eq. (\ref{eqn3}) is the torsion tensor. Finally, the disformation tensor ${%
L^{\gamma }}_{\mu \nu }$ is derived from the non-metricity tensor $Q_{\gamma
\mu \nu }$ as

\begin{equation}
{L^{\gamma }}_{\mu \nu }\equiv \frac{1}{2}g^{\gamma \sigma }\left( Q_{\nu
\mu \sigma }+Q_{\mu \nu \sigma }-Q_{\gamma \mu \nu }\right) .  \label{eqn4}
\end{equation}

In the above equation, the non-metricity tensor $Q_{\gamma \mu \nu }$ is
specific as the (minus) covariant derivative of the metric tensor with
regard to the Weyl-Cartan connection $\Sigma {^{\gamma }}_{\mu \nu }$, i.e. $%
Q_{\gamma \mu \nu }=\nabla _{\gamma }g_{\mu \nu }$, and it can be obtained

\begin{equation}
Q_{\gamma \mu \nu }=-\partial _{\gamma }g_{\mu \nu }+g_{\nu \sigma }\Sigma {%
^{\sigma }}_{\mu \gamma }+g_{\sigma \mu }\Sigma {^{\sigma }}_{\nu \gamma }.
\end{equation}

The connection is presumed to be torsionless and curvatureless within the
current background. It corresponds to the pure coordinate transformation
from the trivial connection mentioned in \cite{ref11}. Thus, for a flat and
torsion-free connection, the connection (\ref{eqn1}) can be parameterized as

\begin{equation}
\Sigma {^{\gamma }}_{\mu \beta }=\frac{\partial x^{\gamma }}{\partial \xi
^{\rho }}\partial _{\mu }\partial _{\beta }\xi ^{\rho }.  \label{eqn6}
\end{equation}

Now, $\xi ^{\gamma }=$ $\xi ^{\gamma }\left( x^{\mu }\right) $ is an
invertible relation. It is always possible to get a coordinate system so
that the connection $\Sigma {^{\gamma }}_{\mu \nu }$ vanish. This condition
is called coincident gauge and has been used in many studies of STEGR \cite%
{ref19} and in this condition the covariant\ derivative $\nabla _{\gamma }$
reduces to the partial derivative $\partial _{\gamma }$. Thus, in the
coincident gauge coordinate, we get

\begin{equation}
Q_{\gamma \mu \nu }=-\partial _{\gamma }g_{\mu \nu }.  \label{eqn7}
\end{equation}

The symmetric teleparallel gravity is a geometric description of gravity
equivalent to GR (STEGR) within coincident gauge coordinates in which $%
\Sigma {^{\gamma }}_{\mu \nu }=0$ and ${C^{\gamma }}_{\mu \nu }=0$, and
consequently from Eq. (\ref{eqn1}) we can conclude that

\begin{equation}
{\Gamma ^{\gamma }}_{\mu \nu }=-{L^{\gamma }}_{\mu \nu }.  \label{eqn8}
\end{equation}

The modified Einstein-Hilbert action in symmetric teleparallel gravity can
be considered as

\begin{equation}
S=\int \left[ \frac{1}{2\kappa }f(Q)+\mathcal{L}_{m}\right] d^{4}x\sqrt{-g},
\label{eqn9}
\end{equation}%
where $\kappa =8\pi G=1$, $f(Q)$ can be expressed as the arbitrary function
of non-metricity scalar $Q$, $g$ is the determinant of the metric tensor $%
g_{\mu \nu }$,\ and $\mathcal{L}_{m}$ is the matter Lagrangian density. Now,
the non-metricity tensor $Q_{\gamma \mu \nu }$ and its traces can be written
as

\begin{equation}
Q_{\gamma \mu \nu }=\nabla _{\gamma }g_{\mu \nu }\,,  \label{eqn10}
\end{equation}%
\begin{equation}
Q_{\gamma }={{Q_{\gamma }}^{\mu }}_{\mu }\,,\qquad \widetilde{Q}_{\gamma }={%
Q^{\mu }}_{\gamma \mu }\,.  \label{eqn11}
\end{equation}

In addition, the superpotential tensor (non-metricity conjugate) can be
expressed as 
\begin{equation}
4{P^{\gamma }}_{\mu \nu }=-{Q^{\gamma }}_{\mu \nu }+2Q_{({\mu ^{^{\gamma }}}{%
\nu })}-Q^{\gamma }g_{\mu \nu }-\widetilde{Q}^{\gamma }g_{\mu \nu }-\delta _{%
{(\gamma ^{^{Q}}}\nu )}^{\gamma }\,,  \label{eqn12}
\end{equation}%
where the trace of the non-metricity tensor can be obtained as 
\begin{equation}
Q=-Q_{\gamma \mu \nu }P^{\gamma \mu \nu }\,.  \label{eqn13}
\end{equation}

Now, the matter energy-momentum tensor is defined as

\begin{equation}
T_{\mu \nu }=-\frac{2}{\sqrt{-g}}\frac{\delta (\sqrt{-g}\mathcal{L}_{m})}{%
\delta g^{\mu \nu }}\,.  \label{eqn14}
\end{equation}

By varying the modified Einstein-Hilbert action (\ref{eqn9}) with respect to
the metric tensor $g_{\mu \nu }$, the gravitational field equations obtained
as 
\begin{widetext}
\begin{equation}
\frac{2}{\sqrt{-g}}\nabla _{\gamma }(\sqrt{-g}f_{Q}P^{\gamma }{}_{\mu \nu })-%
\frac{1}{2}fg_{\mu \nu }+f_{Q}(P_{\nu \rho \sigma }Q_{\mu }{}^{\rho \sigma
}-2P_{\rho \sigma \mu }Q^{\rho \sigma }{}_{\nu })=\kappa T_{\mu \nu }.
\label{eqn15}
\end{equation}%
\end{widetext}
where $f_{Q}=\frac{df}{dQ}$.

\section{Bianchi type-I space-time with field equations}

\label{sec3}

As mentioned in the Introduction, the standard FLRW space-time is isotropic
and homogeneous. Hence, to address the anisotropic nature of the Universe in 
$f\left( Q\right) $ gravity, which manifests as anomalies found in the CMB,
the LRS Bianchi type-I space-time is indeed important because it represents
a spatially homogeneous, but not isotropic. Thus, we consider a Bianchi-type
I space-time in the form 
\begin{equation}
ds^{2}=-dt^{2}+A^{2}(t)dx^{2}+B^{2}(t)(dy^{2}+dz^{2}),  \label{eqn16}
\end{equation}%
where metric potentials $A\left( t\right) $ and $B\left( t\right) $ depend
only on cosmic time $t$. Here, to complete the choice of the anisotropic
type space-time, the equation of state (EoS) parameter of the gravitational
fluid must also be generalized, and from another point of view, to give a
more reasonable model, an anisotropic nature must be presented as described
in \cite{ref30}. \newline

Thus, the energy-momentum tensor for the anisotropic fluid can be expressed
as 
\begin{align}
T_{\nu }^{\mu }& =\text{diag}(-\rho ,p_{x},p_{y},p_{z})\,,  \label{eqn17} \\
& =\text{diag}(-1,\omega _{x},\omega _{y},\omega _{z})\rho ,  \notag \\
& =\text{diag}(-1,\omega ,(\omega +\delta ),(\omega +\delta ))\rho ,  \notag
\end{align}%
where $\rho $ is the energy density of the anisotropic fluid, $p_{x}$, $%
p_{y} $, $p_{z}$ are the pressures and $\omega _{x}$, $\omega _{y}$, $\omega
_{z}$\ are the directional EoS parameters along $x$, $y$ and $z$ coordinates
respectively. The deviation from isotropy is parametrized by setting $\omega
_{x}=\omega $ and then introducing the deviations along $y$ and $z$ axes by
the skewness parameter $\delta $, where $\omega $ and $\delta $ are
functions of cosmic time $t$ \cite{ref27}. \newline
The non-metricity scalar of the anisotropic fluid leads to

\begin{equation}
Q=-2\left( \frac{\dot{B}}{B}\right) ^{2}-4\frac{\dot{A}}{A}\frac{\dot{B}}{B}.
\label{eqn18}
\end{equation}

From the gravitational field equations (\ref{eqn15}), the corresponding
modified Friedmann equations of\ LRS Bianchi type-I space-time (\ref{eqn16})
for the anisotropic fluid of energy-momentum tensor (\ref{eqn17}) can be
written as \cite{ref27}

\begin{widetext}
\begin{equation}
\frac{f}{2}+f_{Q}\left[ 4\frac{\dot{A}}{A}\frac{\dot{B}}{B}+2\left( \frac{%
\dot{B}}{B}\right) ^{2}\right] =\rho ,  \label{eqn19}
\end{equation}

\begin{equation}
\frac{f}{2}-f_{Q}\left[ -2\frac{\dot{A}}{A}\frac{\dot{B}}{B}-2\frac{\ddot{B}%
}{B}-2\left( \frac{\dot{B}}{B}\right) ^{2}\right] +2\frac{\dot{B}}{B}\dot{Q}%
f_{QQ}=-\omega \rho ,  \label{eqn20}
\end{equation}

\begin{equation}
\frac{f}{2}-f_{Q}\left[ -3\frac{\dot{A}}{A}\frac{\dot{B}}{B}-\frac{\ddot{A}}{%
A}-\frac{\ddot{B}}{B}-\left( \frac{\dot{B}}{B}\right) ^{2}\right] +\left( 
\frac{\dot{A}}{A}+\frac{\dot{B}}{B}\right) \dot{Q}f_{QQ}=-(\omega +\delta
)\rho .  \label{eqn21}
\end{equation}
\end{widetext}
where the dot ($\overset{.}{}$) denote derivative with respect to cosmic
time $t$.

The directional Hubble parameters in the direction of the $x$, $y$, and $z$%
-axis, respectively are given by%
\begin{equation}
H_{x}=\frac{\dot{A}}{A},\quad H_{y}=H_{z}=\frac{\dot{B}}{B}.  \label{eqn22}
\end{equation}

The average Hubble parameter, which expresses the volumetric expansion rate
of the Universe is given by 
\begin{equation}
H=\frac{1}{3}\frac{\dot{V}}{V}=\frac{1}{3}\left[ \frac{\dot{A}}{A}+2\frac{%
\dot{B}}{B}\right] ,  \label{eqn23}
\end{equation}%
where the average scale factor and spatial\ volume as 
\begin{equation}
V=a^{3}=AB^{2}.  \label{eqn24}
\end{equation}

The mean anisotropy parameter is given by 
\begin{equation}
\Delta =\frac{1}{3}\sum_{i=1}^{3}\left( \frac{H_{i}-H}{H}\right) ^{2}=\frac{2%
}{9H^{2}}\left( H_{x}-H_{y}\right) ^{2}.  \label{eqn25}
\end{equation}

The expansion scalar $\theta (t)$ and the shear scalar $\sigma (t)$ of the
fluid are defined as follows 
\begin{equation}
\theta (t)=\frac{\dot{A}}{A}+2\frac{\dot{B}}{B},\quad \sigma (t)=\frac{1}{%
\sqrt{3}}\left( \frac{\dot{A}}{A}-\frac{\dot{B}}{B}\right) .  \label{eqn26}
\end{equation}

\bigskip In order to simplify the form of the field equations (\ref{eqn19})-(%
\ref{eqn21}) and write them in terms of the non-metricity scalar $Q$, the
directional Hubble parameters $H_{x}$, $H_{y}$ and average Hubble parameter $%
H$, we use the following relations: $\frac{\partial }{\partial t}\left( 
\frac{\dot{A}}{A}\right) =\frac{\ddot{A}}{A}-\left( \frac{\dot{A}}{A}\right)
^{2}$ and $Q=-2H_{y}^{2}-4H_{x}H_{y}$. The field equations (\ref{eqn19})-(%
\ref{eqn21}) becomes 
\begin{equation}
\frac{f}{2}-Qf_{Q}=\rho ,  \label{eqn27}
\end{equation}%
\begin{equation}
\frac{f}{2}+2\frac{\partial }{\partial t}\left[ H_{y}f_{Q}\right]
+6Hf_{Q}H_{y}=-\omega \rho ,  \label{eqn28}
\end{equation}%
\begin{equation}
\frac{f}{2}+\frac{\partial }{\partial t}\left[ f_{Q}(H_{x}+H_{y})\right]
+3Hf_{Q}\left( H_{x}+H_{y}\right) =-(\omega +\delta )\rho .  \label{eqn29}
\end{equation}

Lastly, here we have three differential equations with six unknowns namely, $%
f$, $H_{x}$, $H_{y}$, $\rho $, $\omega $, and $\delta $. The exact solutions
of these equations are examined in the next section.

\section{Cosmological solutions of field equations}

\label{sec4}

In order to completely solve the field equations, some other constraints
must be added. Although the problems discussed in the introduction above,
the cosmological constant $\Lambda $ in GR is by far the most successful
model among all the proposed alternatives, and thus this motivates us to
examine following linear form of $f(Q)$ gravity model \cite{ref31},

\begin{equation}
f\left( Q\right) =\alpha Q+\beta ,  \label{eqn30}
\end{equation}%
where $\alpha $ and $\beta $ are free model parameters.

Now, using (\ref{eqn30}) and subtracting (\ref{eqn28}) from (\ref{eqn29}),
we get

\begin{equation}
\frac{d}{dt}\left( H_{x}-H_{y}\right) +\left( H_{x}-H_{y}\right) \frac{%
\overset{.}{V}}{V}=-\frac{\delta \rho }{\alpha }.  \label{eqn31}
\end{equation}

This on integrating gives

\begin{equation}
\left( H_{x}-H_{y}\right) =\frac{c}{\alpha V}e^{\int \frac{\delta \rho }{%
\alpha \left( H_{y}-H_{x}\right) }dt},  \label{eqn32}
\end{equation}%
where $c$ is constant of integration.

In order to find the exact solutions to the above equation, we will follow
the work of Adhav \cite{ref32} and Sahni \cite{ref33}, and uses the
condition that

\begin{equation}
\delta =\frac{\alpha }{\rho }\left( H_{y}-H_{x}\right) .  \label{eqn33}
\end{equation}

Using Eq. (\ref{eqn33}) in Eq. (\ref{eqn32}), we obtain this expression

\begin{equation}
\left( H_{x}-H_{y}\right) =\frac{c}{\alpha V}e^{t}.  \label{eqn34}
\end{equation}

The above equation can be written in terms of the metric potentials $A\left(
t\right) $ and $B\left( t\right) $ as $\left( \frac{\overset{.}{A}}{A}-\frac{%
\overset{.}{B}}{B}\right) =\frac{c}{\alpha AB^{2}}e^{t}.$ By looking at this
last equation, we are left with one differential equation and two unknowns,
namely $A\left( t\right) $ and $B\left( t\right) $. Hence, we need a
supplementary constraint to finally solve Eq. (\ref{eqn34}). In this work,
we use the anisotropic relation i.e. the physical condition that the
expansion scalar $\theta \left( t\right) $ is proportional to the shear
scalar $\sigma \left( t\right) $ ($\theta ^{2}\propto \sigma ^{2}$), which
leads to the relation between the metric potentials as%
\begin{equation}
A=B^{n},  \label{eqn35}
\end{equation}%
where $n$ is an arbitrary real number and we think $n\neq 0$, and $1$ for
non-trivial solutions. According to Thorne \cite{Thorne} this physical law is justified on the
basis of the observations of the velocity redshift relation for
extragalactic sources which suggest that the Hubble expansion of the
Universe is isotropic at present time within $30\%$ \cite{Kantowski}. More
exactly, the redshift studies place the limit $\frac{\sigma }{\theta }\leq
0.3$, the ratio of the shear to the expansion scalar in the vicinity of our
galaxy at present time. Collins et al. \cite{Collins} pointed out that the
normal congruence to the homogeneous expansion for spatially homogeneous
metric satisfies the condition $\frac{\sigma }{\theta }=$constant. Bunn et
al. \cite{Bunn} conducted statistical analysis on 4-yr data from CMB and set a
limit for primordial anisotropy to be less than $10^{-3}$ in Planck epoch. Many researchers have used this condition to find exact
solutions of field equations in many backgrounds \cite{ref36, ref37}.

Hence, using the above considerations and condition (\ref{eqn35}), Eq. (\ref%
{eqn34}) takes the form

\begin{equation}
\frac{\overset{.}{B}}{B}-\frac{c}{\alpha \left( n-1\right) B^{n+2}}e^{t}=0,
\label{eqn36}
\end{equation}%
which yields a solution

\begin{equation}
A\left( t\right) =c_{1}^{n}\left[ \frac{\left( n+2\right) e^{t}}{\alpha
\left( n-1\right) }+c_{2}\right] ^{\frac{n}{n+2}},  \label{eqn37}
\end{equation}

\begin{equation}
B\left( t\right) =c_{1}\left[ \frac{\left( n+2\right) e^{t}}{\alpha \left(
n-1\right) }+c_{2}\right] ^{\frac{1}{n+2}},  \label{eqn38}
\end{equation}%
where $c_{1}=c^{\frac{1}{n+2}}$ and $c_{2}$\ both are the constants of
integration. Thus, using Eqs. (\ref{eqn37}) and (\ref{eqn38}), the metric in
Eq. (\ref{eqn16}) takes the form

\begin{widetext}
\begin{equation}
ds^{2}=-dt^{2}+c_{1}^{2n}\left[ \frac{\left( n+2\right) e^{t}}{\alpha \left(
n-1\right) }+c_{2}\right] ^{\frac{2n}{n+2}}dx^{2}+c_{1}^{2}\left[ \frac{%
\left( n+2\right) e^{t}}{\alpha \left( n-1\right) }+c_{2}\right] ^{\frac{2}{%
n+2}}(dy^{2}+dz^{2}).  \label{eqn39}
\end{equation}
\end{widetext}

\section{Evolution of cosmological parameters}

\label{sec5}

In this section, we will discuss some basic physical and geometrical
parameters to validate the cosmological model, such as the spatial volume,
expansion scalar, shear scalar, average Hubble parameter, anisotropic
parameter, energy density, pressure, EoS parameter, skewness parameter, and
deceleration parameter.

Firstly, from Eq. (\ref{eqn18}), the non-metricity scalar becomes

\begin{equation}
Q=-\frac{2(2n+1)e^{2t}}{\left[ c_{2}(n-1)\alpha +(n+2)e^{t}\right] ^{2}}
\label{eqn40}
\end{equation}

The spatial volume of the Universe becomes

\begin{equation}
V=B^{n+2}=c_{1}^{n+2}\left[ \frac{\left( n+2\right) e^{t}}{\alpha \left(
n-1\right) }+c_{2}\right] .  \label{eqn41}
\end{equation}

The expansion scalar and the shear scalar becomes

\begin{widetext}
\begin{equation}
\theta (t)=\frac{(n+2)e^{t}}{c_{2}(n-1)\alpha +(n+2)e^{t}},\quad \sigma (t)=%
\frac{(n-1)e^{t}}{\sqrt{3}\left[ c_{2}(n-1)\alpha +(n+2)e^{t}\right] }.
\label{eqn42}
\end{equation}
\end{widetext}

The average Hubble parameter is obtained as

\begin{equation}
H=\frac{(n+2)e^{t}}{3\left[ c_{2}(n-1)\alpha +(n+2)e^{t}\right] }
\label{eqn43}
\end{equation}

Using Eqs. (\ref{eqn37}), (\ref{eqn38}), and (\ref{eqn43}) in (\ref{eqn25}),
we obtain the anisotropy parameter as follows

\begin{equation}
\Delta =\frac{2(n-1)^{2}}{(n+2)^{2}}.  \label{eqn44}
\end{equation}

From Eqs. (\ref{eqn40}) and (\ref{eqn41}), we observed that the
non-metricity scalar is time-dependent, and the spatial volume of the
Universe is zero in the initial time $t=0$ and increasing function of cosmic
time. Thus, it can be said that in our model the evolution of the Universe
begins with the Big Bang scenario. Also, from Eqs. (\ref{eqn42})-(\ref{eqn43}%
) we can see that the expansion scalar, shear scalar and average Hubble
parameter diverge at $t=0$ and have a finite value at $t\rightarrow \infty $%
. It is also possible to look at the isotropic condition $\frac{\sigma ^{2}}{%
\theta ^{2}}$, as it takes a constant value from the early time to the late
time. Therefore, our model appears that, it does not come close to the
isotropy throughout the evolution of the Universe, and this is confirmed by
Eq. (\ref{eqn44}) where we see that the anisotropic parameter is constant
for our model.

Using Eqs. (\ref{eqn30}) and (\ref{eqn40}) in Eq. (\ref{eqn27}), we obtain
the energy density of the Universe as

\begin{widetext}
\begin{equation}
\rho =\frac{\alpha ^{2}\beta c_{2}^{2}(n-1)^{2}+2\alpha \beta c_{2}\left(
n^{2}+n-2\right) e^{t}+e^{2t}\left[ \alpha (4n+2)+\beta (n+2)^{2}\right] }{2%
\left[ \alpha c_{2}(n-1)+(n+2)e^{t}\right] ^{2}}  \label{eqn45}
\end{equation}

Similarly, using Eqs. (\ref{eqn30}), (\ref{eqn37}), (\ref{eqn38}), (\ref%
{eqn40}), and (\ref{eqn43}) in Eq. (\ref{eqn28}), we obtain the pressure of
the Universe as

\begin{equation}
p=-\frac{\alpha ^{2}\beta c_{2}^{2}(n-1)^{2}+2\alpha c_{2}(n-1)e^{t}\left[
2\alpha +\beta (n+2)\right] +e^{2t}\left[ 6\alpha +\beta (n+2)^{2}\right] }{2%
\left[ \alpha c_{2}(n-1)+(n+2)e^{t}\right] ^{2}}  \label{eqn46}
\end{equation}

Thus, the EoS parameter of the Universe is obtained as

\begin{equation}
\omega =-\frac{\alpha ^{2}\beta c_{2}^{2}(n-1)^{2}+2\alpha c_{2}(n-1)e^{t}%
\left[ 2\alpha +\beta (n+2)\right] +e^{2t}\left[ 6\alpha +\beta (n+2)^{2}%
\right] }{\alpha ^{2}\beta c_{2}^{2}(n-1)^{2}+2\alpha \beta c_{2}\left(
n^{2}+n-2\right) e^{t}+e^{2t}\left[ \alpha (4n+2)+\beta (n+2)^{2}\right] }
\label{eqn47}
\end{equation}

From Eqs. (\ref{eqn33}), (\ref{eqn37}), and (\ref{eqn38}), the skewness
parameter is obtained as

\begin{equation}
\delta =-\frac{2\alpha (n-1)e^{t}\left[ \alpha c_{2}(n-1)+(n+2)e^{t}\right] 
}{\alpha ^{2}\beta c_{2}^{2}(n-1)^{2}+2\alpha \beta c_{2}\left(
n^{2}+n-2\right) e^{t}+e^{2t}\left[ \alpha (4n+2)+\beta (n+2)^{2}\right] }
\label{eqn48}
\end{equation}
\end{widetext}

\subsection{Observational constraints}

In the above discussions, we have described the $f(Q)$ gravity and solved
the field equation. The expressions of the Hubble parameter $H(t)$ in (\ref%
{eqn43}) can be expressed in terms of redshift $z$ as,

\begin{widetext}

\begin{figure}[h]
\centering
\centerline{\includegraphics[scale=0.67]{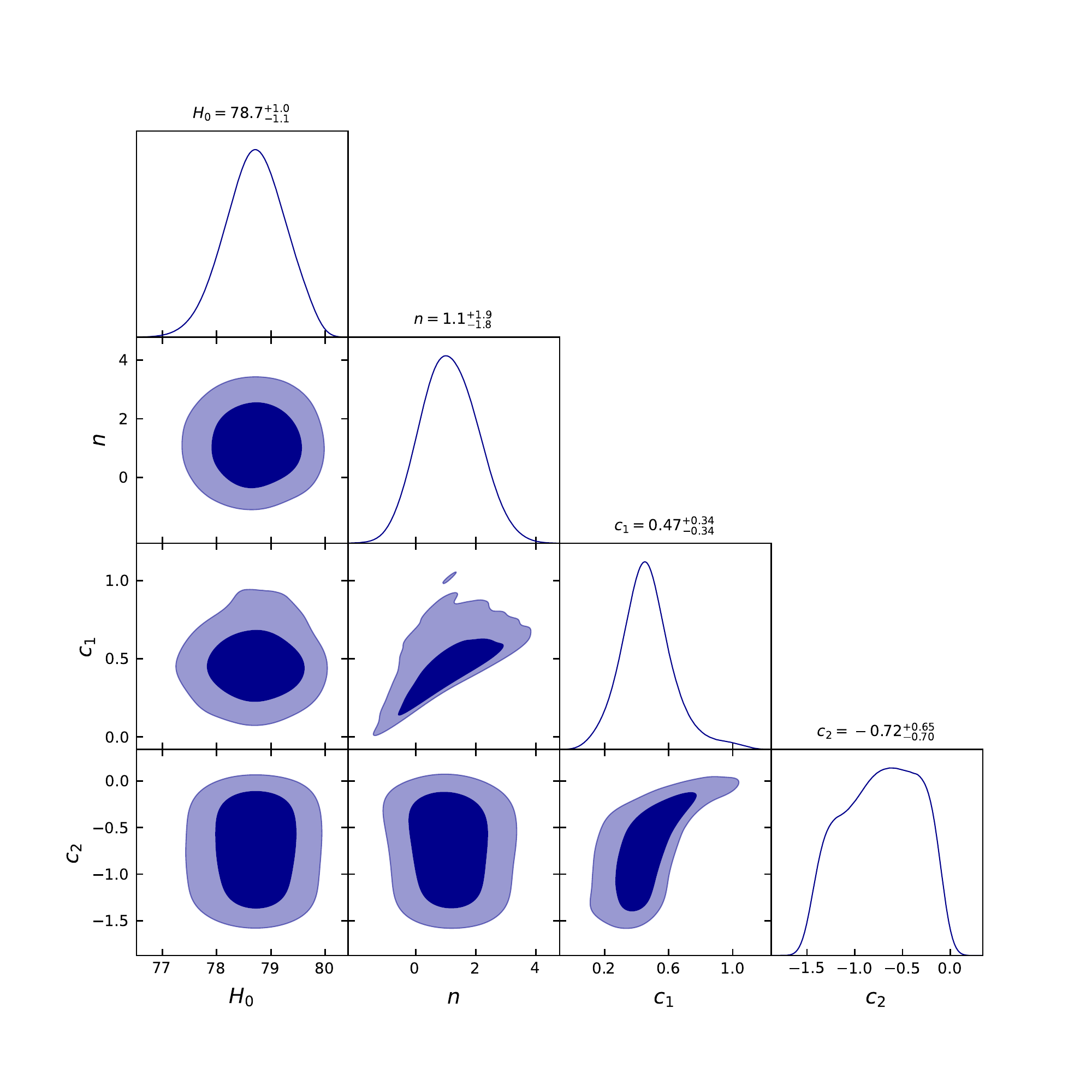}}
\caption{The $1-\protect\sigma $ and $2-\protect\sigma$  likelihood contours
for the model parameters using Hubble datasets.}
\label{contour}
\end{figure}

\begin{figure}[h]
\centering
\centerline{\includegraphics[scale=0.52]{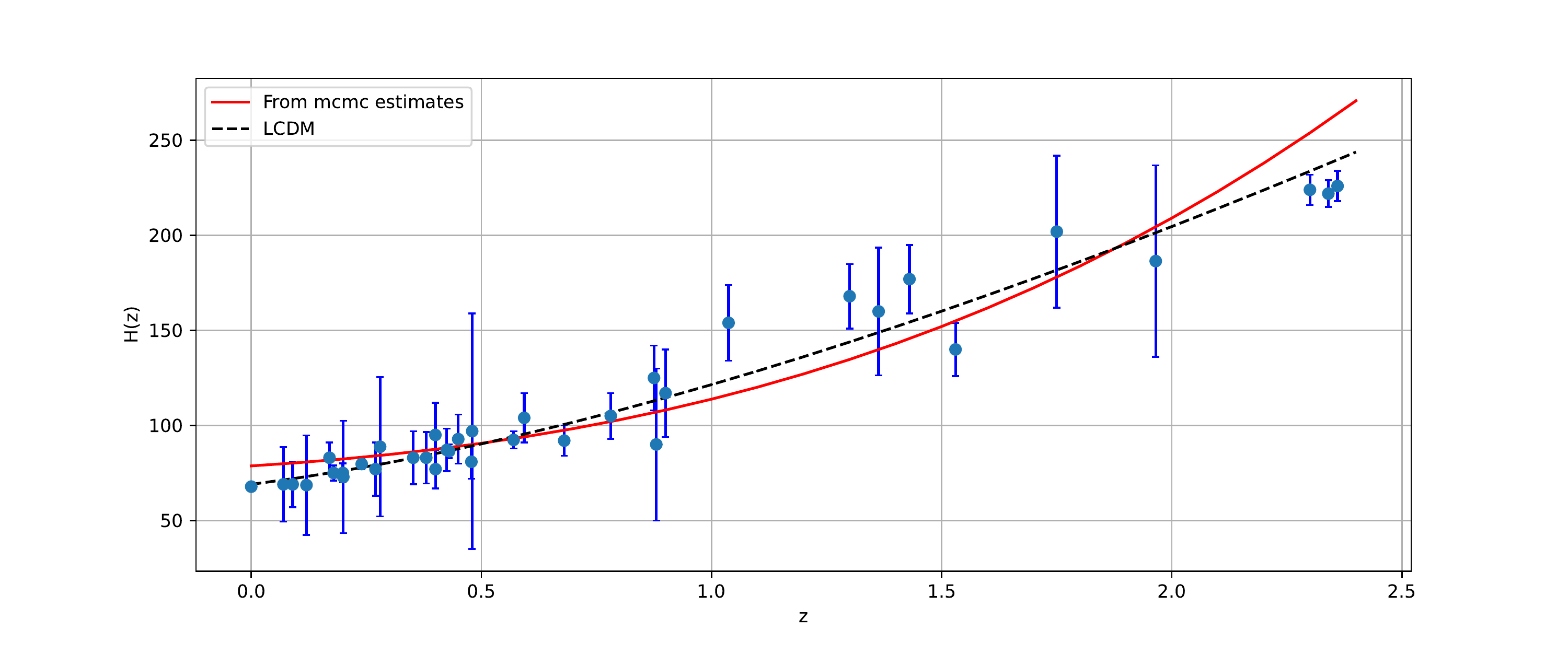}}
\caption{The figure shows the error bar plot of the considered $57$ points of Hubble datasets together with the fitting of Hubble function $H(z)$ vs.
redshift $z$ for our obtained model (red line) compared with that of standard $\Lambda $CDM model (black dashed line).}
\label{hz}
\end{figure}

\end{widetext}

\begin{equation}
H(z)=H_{0}\frac{-1+c_{1}^{n+2}c_{2}\left( 1+z\right) ^{3}}{%
-1+c_{1}^{n+2}c_{2}},  \label{Hz1}
\end{equation}%
where $a\left( t\right) =\left( 1+z\right) ^{-1}$ with $a(t_{0})=a_{0}=1$,
suffix $0$ representing the value of parameter at $t=t_{0}$ and $t_{0}$ is
the present time. The functional form of $H(z)$ contains three model
parameters $c_{1}$, $c_{2}$ and $n$ together with $H_{0}$. In order to
describe the evolotion of some cosmological parameters in our obtained
model, we need to choose some appropriate values of these model parameters.
So, we consider here the Observational Hubble Datasets (OHD) to get some
best fit values of these model parameters. We have used recently compiled $57
$ data points from OHD as in the reference \cite{sharov}, which is used in
several papers. Scipy optimization technique from Python library is used
here together with the consideration of a Gaussian prior with a fixed $%
\sigma =1.0$ as the dispersion using Python's emcee package. The results are
shown in the contour plots (two-dimensional) with $1-\sigma $ and $2-\sigma $
errors. The Chi-square function for our analysis is given by,

\begin{equation}
\chi _{H}^{2}(c_{1},c_{2},n)=\sum\limits_{i=1}^{57}\frac{%
[H_{th}(z_{i},c_{1},c_{2},n)-H_{obs}(z_{i})]^{2}}{\sigma _{H(z_{i})}^{2}},
\label{chihz}
\end{equation}%
where $H_{obs}$ is the observed value of the Hubble parameter and $H_{th}$
is its theorised value and the symbol $\sigma _{H(z_{i})}$ is the standard
error in the observed value of the $H(z)$. With the above set up, we have
found the best fit values of the model parameters for the observational
Hubble dataets as $c_{1}=0.191_{-0.093}^{+0.093}$, $%
c_{2}=1.21_{-0.13}^{+0.13}$, $n=1.21_{-0.13}^{+0.13}$. The result is shown
in Fig. (\ref{contour}) as a two dimensional contour plots with $1-\sigma $
and $2-\sigma $ errors.

Additionally, we observed our derived model has nice fit to the
aforementioned Hubble datasets. The error bars for the considered datasets
and the $\Lambda $CDM model (with $\Omega _{\Lambda 0}=0.7$ and $\Omega
_{m0}=0.3$) are also plotted along with our model for comparision. This is
displayed in Fig. (\ref{hz}),

\subsection{Physical interpretation of some cosmological parameters of the model}

In cosmology, the deceleration parameter $q$ is a measure of the variation
in the expansion of the Universe, if $q<0$ the Universe is in a phase of
accelerated expansion and if $q>0$ the Universe is in a phase of decelerated
expansion and is defined as $q=-1+\frac{d}{dt}\left( \frac{1}{H}\right) $.
For the model under discussion, the deceleration parameter is obtained as,

\begin{equation}
q=\frac{e^{-t}\left[ (n+2)\left( -e^{t}\right) -3c_{2}(n-1)\alpha \right] }{%
n+2},  \label{eqn49}
\end{equation}%
which can be written in terms of redshift $z$ as,

\begin{equation}
q(z)=-1+\frac{3c_{1}^{n+2}c_{2}\left( 1+z\right) ^{3}}{-1+c_{1}^{n+2}c_{2}%
\left( 1+z\right) ^{3}}.  \label{qz1}
\end{equation}

\begin{figure}[h]
\centerline{\includegraphics[scale=0.65]{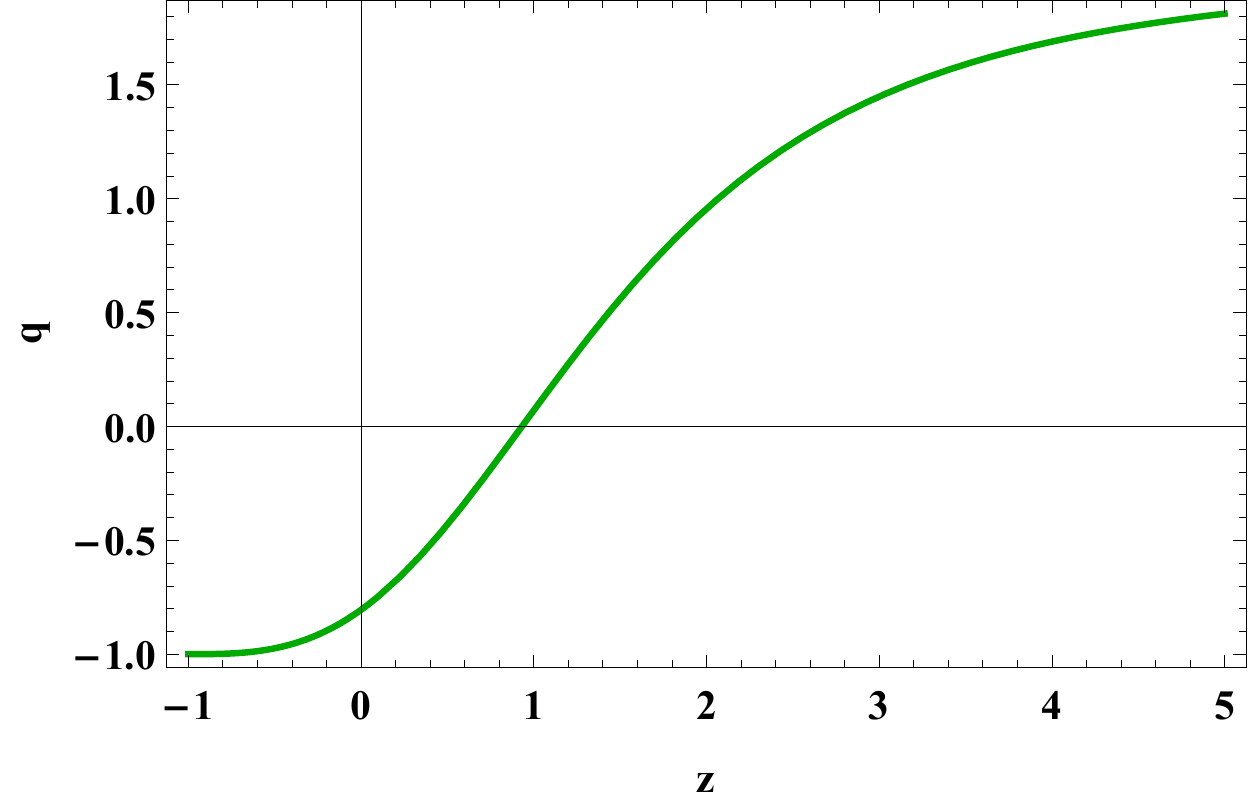}}
\caption{$q$ versus redshift $z$corresponding to the values of the
parameters constrained by the Hubble datasets.}
\label{fig5}
\end{figure}

The plot for the deceleration parameter in Fig. (\ref{fig5}) exhibits a
phase transition from early deceleration to the current acceleration of the
Universe with current value corresponding to the observational Hubble datasets $q_{0}\sim -0.7804$.

\begin{figure}[h]
\centerline{\includegraphics[scale=0.65]{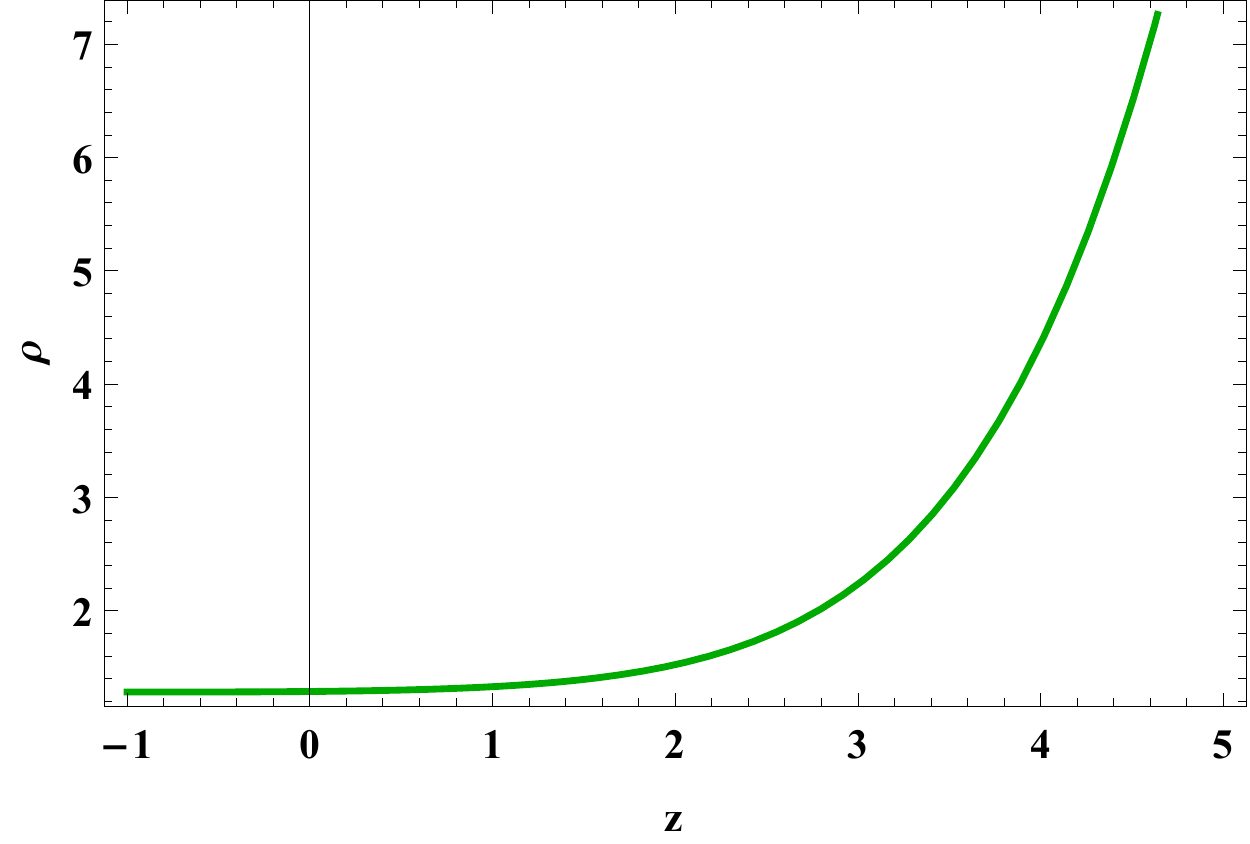}}
\caption{$\protect\rho $ versus redshift $z$ corresponding to the values of
the parameters constrained by the Hubble datasets.}
\label{fig1}
\end{figure}

\begin{figure}[h]
\centerline{\includegraphics[scale=0.65]{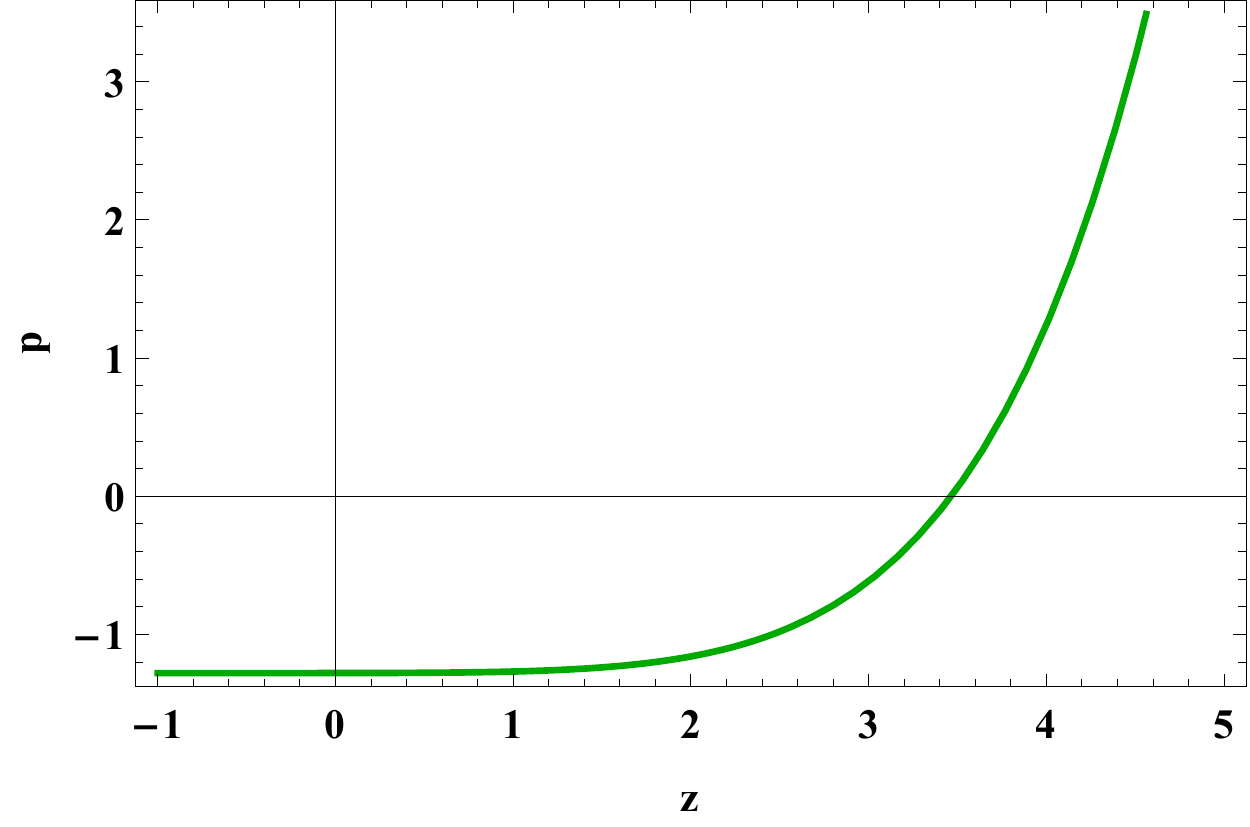}}
\caption{$p$ versus redshift $z$ corresponding to the values of the
parameters constrained by the Hubble datasets.}
\label{fig2}
\end{figure}

\begin{figure}[h]
\centerline{\includegraphics[scale=0.65]{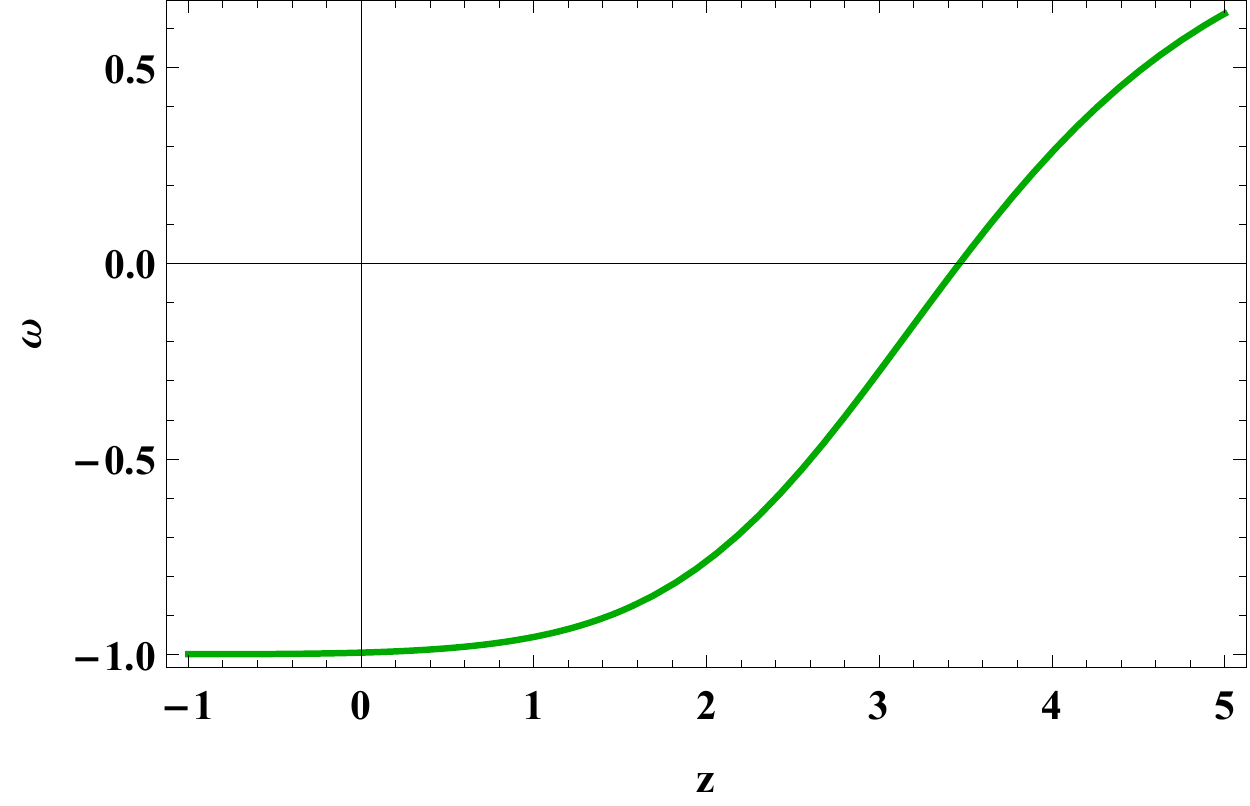}}
\caption{$\protect\omega $ versus redshift $z$ corresponding to the values
of the parameters constrained by the Hubble datasets.}
\label{fig3}
\end{figure}

\begin{figure}[h]
\centerline{\includegraphics[scale=0.66]{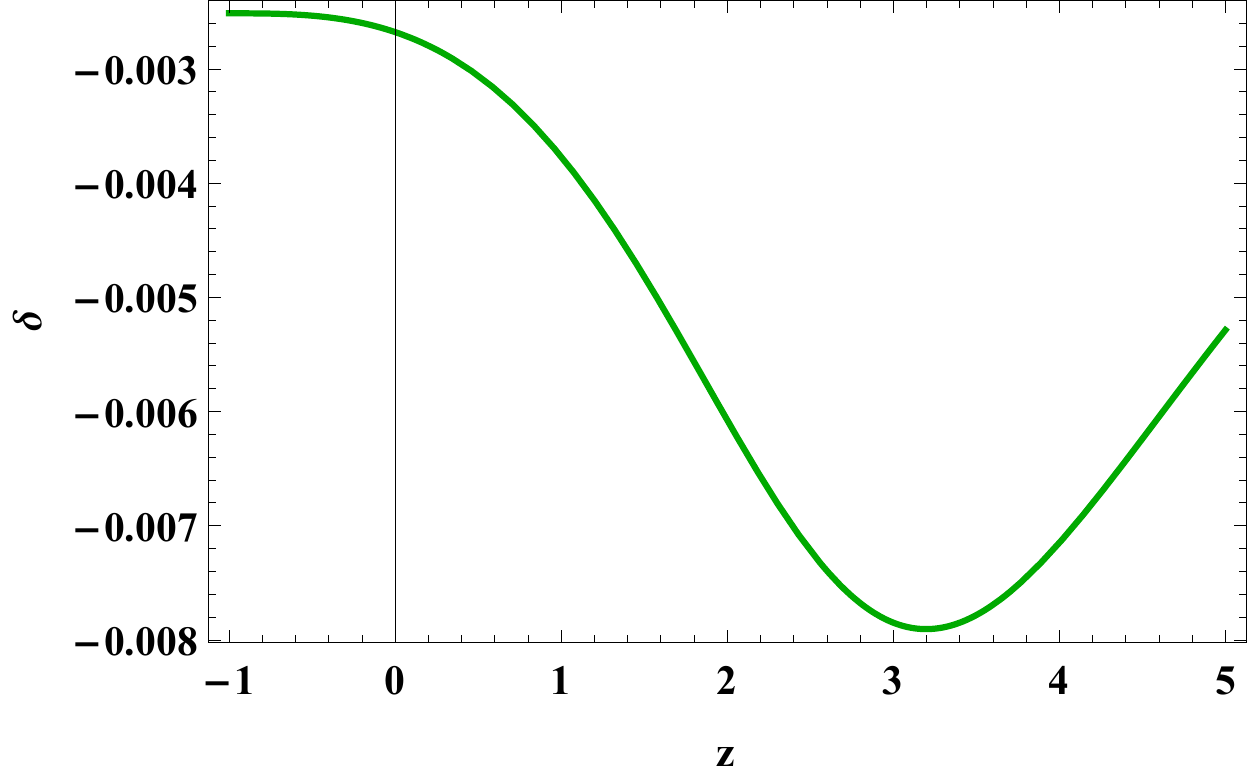}}
\caption{$\protect\delta$ versus redshift $z$ corresponding to the values of
the parameters constrained by the Hubble datasets.}
\label{fig4}
\end{figure}

From Fig. (\ref{fig1}), it is clear that the energy density of the Universe
is an increasing function of cosmic redshift $z$ and also it remains
positive for all redshift values. It starts with large positive values and approaches zero in the
future i.e. $z\rightarrow -1$. Fig. (\ref{fig2}) indicates that the pressure
of the Universe is also an increasing function of cosmic redshift $z$, which
starts with large positive values and then becomes negative in the present
period ($z=0$) and in the future, the negative pressure pushes the Universe
to the phase of acceleration, as indicated by astronomical observations. The EoS parameter is a relationship between energy density and pressure, and
helps us determine the phases the Universe has gone through. The matter
phase at $\omega =0$. Next, $\omega =\frac{1}{3}$ exhibit the
radiation-dominated phase, while $\omega =-1$ corresponds to the $\Lambda $%
CDM model. In addition, the acceleration phase of the Universe is described
at $\omega <-\frac{1}{3}$ which includes the quintessence ($-1<\omega \leq -%
\frac{1}{3}$) and phantom model ($\omega <-1$).
Moreover, the EoS parameter presented in Fig. (\ref{fig3}) indicates that
the anisotropic fluid behaves like the $\Lambda $CDM model \cite{ref38}. The current value of EoS parameter corresponding to the observational Hubble datasets is $\omega _{0}\sim -1$. In Fig. (\ref{fig4}) we see that
the skewness parameter evolves in the range of negative values and tends
towards values close to zero in the future, which confirms the previous
discussion that our model remains under anisotropic behavior throughout the
expansion of the Universe. Thus, this anisotropic cosmological $f(Q)$ model
simulates the standard $\Lambda $CDM model.

\section{Concluding remarks}

\label{sec6}

In this paper, we investigated the homogeneous and anisotropic LRS Bianchi-I
space-time in the framework of $f\left( Q\right) $ modified gravity, where
the non-metricity $Q$ is the basis of gravitational interactions with zero
curvature and torsion. The physical motivation for exploring the anisotropic
Universe is the small deviations from the isotropy observed by the nine-year
Wilkinson Microwave Anisotropy probe (WMAP) \cite{ref39}, which could yield
more realistic results, especially with the $f(Q)$ modified theory of
gravity. First, we briefly presented the mathematical formalism of the
theory, then we derived the field equations for the LRS Bianchi-I space-time
for the content of the Universe in the form of a perfect anisotropic fluid
as in references \cite{ref27, ref28}. To get the exact solutions
and study dark energy in $f(Q)$ gravity, motivated by the
cosmological constant ($\Lambda $), we have considered the following linear
model $f\left( Q\right) =\alpha Q+\beta $, where $\alpha $ and $\beta $ are
free model parameters. Further, to complete the solutions we used the
assumption that the scalar expansion $\theta \left( t\right) $ is
proportional to the shear scalar $\sigma \left( t\right) $, which leads to
the relation between the metric potentials in the form $A=B^{n}$, where $n$
is an arbitrary constant. We obtained the best fit values of the model parameters by using the observational
Hubble datasets of 57 data points. The obtained best fit values are $c_{1}=0.191_{-0.093}^{+0.093}$, $%
c_{2}=1.21_{-0.13}^{+0.13}$, $n=1.21_{-0.13}^{+0.13}$.

Under these considerations, we found the complete solutions to the field
equations and we have investigated the behavior of some cosmological parameters such
as: The spatial volume of the Universe is zero in the initial time $t=0$,
which suggests that the evolution of the Universe begins with the Big Bang
scenario and thus the model has a point type singularity \cite{ref40}. The
expansion scalar, shear scalar, and average Hubble parameter diverges at $%
t=0 $ and become a finite value at $t\rightarrow \infty $. To test the
anisotropy of the model, we have studied the behavior of the anisotropic
parameter and found that it takes a constant value throughout the expansion
of the Universe. Further, for physical properties, we have discussed the
behavior of energy density $\rho $, pressure $p$, equation of state (EoS)
parameter $\omega $, and skewness parameter $\delta $ with the help of Figs.
(\ref{fig1})-(\ref{fig4}). We have found the positive energy density and
negative pressure, which results in the EoS parameter behaving like the
standard $\Lambda $CDM model and its current value corresponding to the observational Hubble datasets is $\omega _{0}\sim -1$.
Thus, this value is consistent with the observational constraints on the
dark energy EoS $\omega $ such as $\omega _{0}=-1.03\pm 0.03$ \cite{ref24},
which suggests its value should be highly close to -1. In our model, we have
discussed the behavior of skewness parameter which is an effective tool for
checking whether the model is anisotropic or not, because in the case of an
isotropic Universe, $\delta =0$, and we have found that it changes in the
range of negative values and tends towards values close to zero in the
future, which confirms that our model remains under anisotropic behavior
throughout the expansion of the Universe. In addition, we observed that for
all values of $n$ the deceleration parameter exhibits a phase transition
from early deceleration to the current acceleration of the Universe with
current value corresponding to the observational Hubble datasets $q_{0}\sim -0.7804$. Finally, it can be said that this type of
result agrees well with the accelerating scenario of the Universe.

\section*{Acknowledgments}

We are very much grateful to the honorary referee and the editor for the
illuminating suggestions that have significantly improved our work in terms
of research quality and presentation.

\textbf{Data availability} There are no new data associated with this article%
\newline

\textbf{Declaration of competing interest} The authors declare that they
have no known competing financial interests or personal relationships that
could have appeared to influence the work reported in this paper.\newline


\end{document}